\documentclass[twocolumn,showpacs,preprintnumbers,amsmath,amssymb,aps,psfig,epsfig]{revtex4}
\usepackage{epsfig}

\newcommand{\smallstep}{\vspace{.06em}}
\def\di{\displaystyle}

\def\bg{\begin{eqnarray}\begin{array}{rcl}\displaystyle}
\def\eg{\end{array} &\di    &\di   \end{eqnarray}}
\def\bm#1{\begin{eqnarray}\begin{array}{#1}\di}
\def\bmo#1{\begin{eqnarray*}\begin{array}{#1}\di}
\def\bml#1#2{\begin{eqnarray}\begin{array}{#1}\label{#2}\di}
\def\bgo{\begin{eqnarray*}\begin{array}{rcl}\displaystyle}
\def\ego{\end{array} &\di    &\di \nonumber  \end{eqnarray*}}

\def\btensor#1#2{\renew\left#1\begin{array}{#2}\di}
\def\brtensor#1#2#3{\ren#3\left#1\begin{array}{#2}}
\def\botensor#1#2{\renew\left#1\begin{array}{#2}}
\def\etensor#1{\end{array}\right#1}

\def\eq#1{(\ref{#1})}
\def\Eq#1{Eq.~(\ref{#1})}

\def\tr{{\rm tr}}
\def\Tr{{\rm Tr}}

\def\id{1\!\mbox{l}}

\def\s0#1#2{\mbox{\small{$ \frac{#1}{#2} $}}}
\def\0#1#2{\frac{#1}{#2}}

\def\ren#1{\renewcommand{\arraystretch}{#1}}
\def\rene{\renewcommand{\arraystretch}{1.9}}
\def\renew{\renewcommand{\arraystretch}{1}}
\rene

\begin{document}

\title{Infrared behaviour and fixed points in Landau gauge QCD}
 
\vspace{1.5 true cm}
 
\author{Jan M. Pawlowski${}^a$, Daniel F. Litim${}^b$, 
Sergei Nedelko${}^{a,c}$, 
Lorenz von Smekal${}^a$\\[-2ex] \ }

\affiliation{\mbox{\it ${}^a$ Institut f\"ur Theoretische Physik III, 
Universit\"at Erlangen, 
Staudtstra\ss e 7, 91058 Erlangen, Germany}\\ 
\mbox{\it ${}^b$Theory Division, CERN, CH-1211 Geneva 23 and 
SPA, U. Southampton, Southampton SO17 1BJ, U.K.}\\ 
${}^c$BLTP, JINR, 141980 Dubna, Russia}
\preprint{FAU-TP3-03-12, CERN-TH-2003-300, SHEP-03-33}

\begin{abstract}
  {We investigate the infrared behaviour of gluon and ghost
    propagators in Landau gauge QCD by means of an exact
    renormalisation group equation. We explain how, in general, the
    infrared momentum structure of Green functions can be extracted
    within this approach. An optimisation procedure is devised to  
    remove residual regulator dependences. In Landau gauge QCD
    this framework is used to determine the infrared leading terms of
    the propagators. The results support the Kugo-Ojima confinement
    scenario.  Possible extensions are discussed.  }
\end{abstract}


\maketitle

\pagestyle{plain}
\setcounter{page}{1}

Charged particles have long-range nature, whereas localised objects
are neutral.  For local quantum field theories one derives more
generally that every gauge-invariant localised state is singlet with
respect to the unbroken global charges of the gauge symmetry.  Thus,
in QCD without Higgs mechanism any localised physical state must be
colourless. This extends to all physical states only if a mass gap is
present. Then, colour-electric charge superselection sectors cannot
arise, and every gauge-invariant state must also be a colour singlet.
These are the signatures of confinement.\smallstep

Within covariant linear gauges, the necessary conditions for
confinement were formulated more than twenty years ago by Kugo and
Ojima \cite{Kug79}.  They express the necessity of a mass gap and the
conditions for avoiding the Higgs mechanism in a covariant continuum
formulation of QCD in terms of local fields, {\it i.e.}, for the local
measurement of (colour) quantum numbers.  Both conditions are encoded
in the critical infrared exponents of gluonic and ghost correlations
in Landau gauge QCD \cite{Alkofer:2000wg}. For the mass gap, the
massless transverse gluon states of perturbation theory must be
screened nonperturbatively, leading to an infrared suppression for
the gluon propagator. In turn, the global gauge charges entail the
infrared enhancement of the ghost correlations leading to their
infrared dominance.  \smallstep

The infrared behaviour just described has been found in solutions of
truncated Dyson-Schwinger equations \cite{Sme97}. Since then, the same
qualitative behaviour has emerged in a variety of nonperturbative
approaches including an increasing number of lattice simulations
\cite{Cucchieri:1997fy} and investigations based on Fokker-Planck type
diffusion equation of stochastic quantisation \cite{Zwa02}.
Quantitatively, the latter leads to the same infrared exponents as the
Dyson-Schwinger equations, which reflects an equivalence between the
two formulations under certain conditions \cite{Zwa02,Lerche:2002ep}.
Interesting in their own right, such interrelations are especially
useful for finding ways to go beyond the particular approximations
employed in each case.  \smallstep

Here we present an investigation of the infrared sector of Landau
gauge QCD by means of a nonperturbative flow 
equation \cite{Wegner,Wetterich:yh}. Central to the approach is the
effective action $\Gamma_k$, where quantum fluctuations with
momenta $p^2>k^2$ are already integrated out, and $k$ denotes an
infrared cut-off scale. The variation with $k$ leads to a flow
equation for $\Gamma_k$, thereby interpolating between the classical
action in the ultraviolet and the full quantum effective action in the
infrared where the cut-off is removed. Moreover, it has been shown
that truncations can be systematically improved by means of an
optimisation procedure \cite{Litim:2001up}. The formalism has its particular
merits in gauge theories as it is amiable to approximations being
non-local in momenta and fields, precisely the behaviour we expect in
QCD. So far it has been applied to Landau gauge QCD for a
determination of the heavy quark effective potential
\cite{Ellwanger:1996qf} and effective quark interactions above the
confinement scale \cite{Bergerhoff:1997cv}; for a brief review and
further applications in Yang-Mills theories see \cite{Litim:1998nf} and
\cite{Pawlowski:1996ch} respectively.\smallstep

Conceptually, the flow comprises a system of coupled differential
equations for all Green functions which only involves dressed vertices
and propagators. Hence, the correct RG-scaling is displayed by the
full flow and truncations with the correct symmetry properties. This
is one of the main contradistinctions to Dyson-Schwinger equations or
stochastic quantisation which involve both bare and dressed Green
functions.  This makes it often difficult in the latter two approaches
to implement the correct RG-scaling behaviour within truncations.
\smallstep

In this letter we explain how the full
infrared momentum behaviour of Green functions is extracted within a
fixed point regime.  In Landau gauge QCD, this allows for a simple
determination of infrared coefficients for gluon and ghost propagator
as well as the running coupling.  Results for general regulators are
presented and interpreted in the context of an optimisation of the
flow.  The systematic extension of the truncation is
discussed. \smallstep

We begin by introducing the classical gauge fixed action for QCD in a
general covariant gauge in four dimensions. Including also the ghost
action, it is given by
\begin{eqnarray}
S_{\rm cl}=
\s012 \int \tr\, F^2
+\s0{1}{2\xi}\int\,(D_\mu A^\mu)^2 
+\int\,\bar C\cdot \partial_\mu D_\mu\cdot C\, . 
\label{sclassical} 
\end{eqnarray}
\smallstep 
The flow equation approach relies on a momentum cut-off for the propagating 
degrees of freedom. Here, the momentum regularisation for 
 both, the gluon and the ghost, is achieved by 
$S_{\rm cl} \to S_{\rm cl}+\Delta S_k$ with 
\begin{eqnarray}\label{dSk}
\Delta S_k=
\s012 \int \, A^\mu_a\, R_{\mu\nu}^{ab} \, A^\nu_b 
+\int \, \bar C_a\, R^{ab}\,C_b\,.
\end{eqnarray} 
The functions $R$ implement the infrared momentum cut-off at the
momentum scale $p^2\approx k^2$. The scale-dependence of the
regularisation in \eq{dSk} induces a flow equation for the
effective action. With $\phi=(A,C,\bar C)$ and $t=\ln k$, the flow
equation takes the explicit form
\begin{eqnarray}\nonumber 
\lefteqn{{\partial_t} \Gamma_k[\phi]=  
\frac{1}{2} \int d^4 p \ G_{ab}^{\mu\nu}[\phi](p,p)
\ {\partial_t} R_{\mu\nu}^{ba}(p)}\hspace{1.3cm}\\ 
&& -
\int d^4 p \ G^{ab}[\phi](p,p)
\ {\partial_t} R^{ba}(p)\,,
\label{flow}\end{eqnarray} 
where $ G[\phi](p,q)= \bigl(\Gamma_k^{(2)}+R \bigr)^{-1}(p,q) $
denotes the full regularised propagator.  The flow is finite in both,
the infrared and the ultraviolet, by construction.  Effectively, the
momentum integration in \eq{flow} only receives contributions for
momenta in the vicinity of $p^2\approx k^2$. Consequently 
it has a remarkable numerical stability.  The flow solely depends on
dressed vertices and propagators leading to consistent RG-scaling on
either side of \eq{flow}.  The flow equations for propagators and
vertices are obtained from \eq{flow} by functional
derivatives.\smallstep

To verify the Kugo-Ojima confinement criterion we have to determine
the momentum behaviour of the propagators in the deep infrared region
\begin{equation}
\label{validity}
k^2\ll 
p^2\ll 
\Lambda^2_{\rm QCD}\,.
\end{equation}
In this domain, quantum fluctuations are already integrated out and
the {\it physical} quantities as well as general vertex functions will
show no $k$ dependence. Moreover, \eq{validity} entails that we are
already deep in the infrared regime. No new physics arises and the
flow equation \eq{flow} only constitutes the shifting of the transition 
regime at momenta about $k^2$ towards smaller scales. 
Accordingly, Green functions $\Gamma^{(n)}_k\equiv\delta^{(n)}\Gamma_k
/ \delta\phi_1\cdots\delta\phi_n$ can be parameterised as
\begin{eqnarray}\label{green}
\Gamma_k^{(n)}(p_i^2\ll\Lambda^2_{\rm QCD})
=z_n\,\hat\Gamma^{(n)}(p_i,p^2_i/k^2)\,,  
\end{eqnarray}
where $p_i$ stands for the momenta $p_1,...,p_n$.  Here $z_n$ are
possibly $k$-dependent prefactors accounting for an RG-scaling that
leaves the full action invariant. This reduces the remaining
$k$-dependence in $\hat \Gamma^{(n)}$ to dimensionless ratios
$x=p^2/k^2$. Hence, the interpolation between the physical infrared
regime \eq{validity} and the regularised behaviour for $p^2\ll k^2$ is
$k$-independent. This implies a fixed point behaviour. So far, fixed
point investigations within the exact renormalisation group have been
put forward within derivative expansions. Here we provide a new
truncation scheme for accessing the full momentum dependence of Green
functions in the infrared limit.  \smallstep

Taking advantage of their simplified structure in the physical regime
\eq{validity}, we introduce the following parametrisation for the
gluon and ghost two point function
\begin{eqnarray}\nonumber 
\Gamma_{k,A}^{(2)}(p^2)&=& z_A\cdot Z_A(x)\cdot
p^2\cdot \Pi(p)\cdot \id +{\rm longitudinal}\,,\\
\Gamma_{k,C}^{(2)}(p^2)&=& z_C\cdot Z_C(x)\cdot p^2\cdot \id\,, 
\label{parG2}\end{eqnarray} 
where
\begin{eqnarray}
\nonumber
Z(x)= x^{\kappa}\bigl(1+\delta Z(x)\bigr)\,. 
\end{eqnarray}
Here $\Pi_{\mu\nu}(p)=\delta_{\mu\nu}-p_\mu p_\nu/p^2$,
$\id_{ab}=\delta_{ab}$ and $Z,\,\delta Z,\,\kappa$ without subscripts
stand for $Z_{A/C},\,\delta Z_{A/C},\,\kappa_{A/C}$. The longitudinal
part of the gluon two point function in \eq{parG2} is not given
here as it does not contribute to the Landau gauge flows
considered below.  The momentum dependence of \eq{parG2} in the
asymptotic regimes $x\gg 1$ and $x\ll 1$ fixes the functions $\delta
Z$ in both regimes: the prefactor $x^\kappa$ encodes the leading
behaviour for the momentum regime \eq{validity} where $1\ll x\ll
\Lambda^2_{\rm QCD}/k^2$, hence $\delta Z$ tends to zero. 
The $k$-independence of \eq{parG2} at $k=0$
implies $\partial_t \kappa=0$ and $\partial_t z=2 \kappa\, z$. In the
limit of small $x$, we recover the trivial momentum behaviour
$\Gamma_k^{(2)}(p^2)= m_k^2+z_1\, p^2+O(p^3)$ as a consequence of the
cut-off. Accordingly the functions $\delta Z$ have the limits
\begin{eqnarray}  \nonumber
\delta Z(x\to \infty)&\to& 0\,,\\
\delta Z(x\to 0)&\to &
-1+\mu\, x^{-(1+\kappa)}+O(x^{-\kappa})\,.
 \label{dZconstraints} 
\end{eqnarray}  
We emphasise that the limit $x\to \infty$ implies $k\to 0$ as we demand
$x\ll \Lambda^2_{\rm QCD}/k^2$. In the intermediate region $x \approx
1$ the propagators show a regulator-dependent interpolation between
the physical infrared behaviour and the trivial cut-off regime. Also
\begin{eqnarray}\label{trivextra} 
\partial_t \, \delta Z(x)=-2x\partial_x\, \delta Z
\end{eqnarray} 
reflects that the dependence of $\delta Z$ on $k$ stems from the
interpolation property while its shape is independent of it.  The
vertices $\Gamma_k^{(n)}$ in the present truncation are schematically
given by
\begin{eqnarray}\label{vertices} 
\Gamma_k^{(n)}= z_n\, S^{(n)}_{\rm cl}
\end{eqnarray} 
for $ n\geq 3$. In the above parametrisation, the
non-re\-nor\-malisation of the ghost-gluon vertex at $k=0$ leads to
\begin{eqnarray}\label{non-ren}
\kappa_A=-2 \kappa_C\,,\quad\quad  \alpha_s= \0{g^2}{4 \pi}\, 
\0{1}{ z_A\, z_C^2}\;. 
\end{eqnarray} 
It follows from \eq{non-ren} and
$\partial_t z=2 \kappa\,z$ that the strong coupling constant achieves
a fixed point in the infrared limit, $\partial_t \alpha_s=0$.  In
particular, the $z$-factors for the three- and four-gluon vertices are
proportional to $\alpha_s^{1/2} z_A^{3/2}$ and $ \alpha_s z_A^2$,
respectively. Moreover it follows within the truncation \eq{vertices}
that no mass term for the ghost is present corresponding to
$\mu_C=0$ in \eq{dZconstraints}. 
In summary, the truncation \eq{parG2},\eq{vertices} with
the properties \eq{dZconstraints},\eq{non-ren} satisfy the RG
properties of Landau gauge QCD as well as the truncated 
Slavnov-Taylor identities valid in the presence of the regulator. \smallstep

We proceed with determining the strong coupling fixed point $\alpha_s$
and the infrared exponents $\kappa$ from \eq{flow} in the truncations
put down in \eq{parG2} .  We fix the tensor structure of the
regulators as $R_{\mu\nu}^{ab}=R_A \delta_{\mu\nu}\delta^{ab}$ and
$R^{ab}=R_C\delta^{ab}$, leaving the scalar part $R= z\, p^2\, r(x)$
at our disposal. The integrated flow leads to an integral equation for
$\delta Z$,  
\begin{eqnarray}\label{genfixed}
\delta Z_{A/C}(x)& = & \0{\alpha_s}{\pi^2} N   \int_x^\infty 
\0{d x'}{{x'}^{2+\kappa}} \, f_{A/C}(x')\,, 
\end{eqnarray} 
where
\begin{eqnarray*}
\label{genfA} 
f_A(x)&=&-\013 
\int_0^\infty dy\, y\int_{-1}^{1}dt\,\0{y}{u}\frac{(1-t^2)^{3/2}
{\cal T}_C(y)}{r_C(u)+Z_C(u)}\,,
\\
f_C(x)&=&\frac{x}{2}\int_0^\infty dy\,y \int_{-1}^{1}dt(1-t^2)^{3/2}
\\
&&\times\0{1}{u}
\left(\0{y}{u}
\frac{{\cal T}_C(y)}{r_A(u)+Z_A(u)}+\frac{{\cal T}_A(y)}{r_C(u)+Z_C(u)}
\right). 
\end{eqnarray*}
and $u=x+y+2t\sqrt{xy}$. We have also introduced the 
abbreviation $2 {\cal T}=k^2 G (\partial_t R) G $, which reads explicitly
\begin{eqnarray}\nonumber
{\cal T}(y)=\0{1}{y}\frac{\kappa\, r(y)-y\, r'(y)}{(r(y)+ Z(y))^2 }\,.
\end{eqnarray}  
In \eq{genfixed}, the integration over $k$ was traded for an
integration over $x$, using $z=\hat z\, x^{-\kappa} (\Lambda^2_{\rm
QCD}/p^2)^\kappa$ with $\partial_t \hat z=0$. The purely gluonic
diagrams in \eq{flow} are suppressed by positive powers of
$p^2/\Lambda^2_{\rm QCD}$ in the fixed point regime and do not
contribute to \eq{genfixed}.  \smallstep

For the sharp cut-off the kernel ${\cal T}(x)$ simplifies and becomes
${\cal T}(x)=\delta(x-1)/(z\,Z(1))$. Then the $t$ and $y$ integrals in
\eq{genfixed} can be performed analytically leading to a lengthy sum
of hyper-geometric functions. For generic smooth cut-off, $\kappa$ and
$\alpha_s$ are computed from \eq{genfixed} in the limit $x\to 0$ as
follows. On the right hand side of \eq{genfixed}, the integrands
$f_{A/C}$ are suppressed for small momenta $x$ and $y$ due to ${\cal
  T}_{A/C}$. This implies that $\delta Z$ can only contribute for
sufficiently large arguments where it tends to zero due to
\eq{dZconstraints}. Therefore we can safely neglect $\delta Z$ under
the momentum trace on the right hand side of \eq{genfixed} in a first
step. Then, we proceed to fully consistent solutions $\delta Z$ of
\eq{genfixed} by iteration. On the left hand side of the gluon
equation the term $\delta Z_A(x\to 0)$ is determined by
\eq{dZconstraints}. In the ghost equation, we have to carefully
identify the terms. For $\kappa_C>0$ we read-off from
\eq{dZconstraints} that $\delta Z_C(x\to 0)$ diverges. However,
$\Gamma_{k,C}^{(2)}(p^2)=z_C x^{\kappa_C} (1+\delta Z_C(x))$ has a
Taylor expansion about $x=0$ due to the IR regularisation. Hence the
coefficient of the constant term in $\delta Z_C$ is $-1$.  \smallstep

The integral equation \eq{genfixed} can be solved numerically for
$\kappa$ and $\alpha_s$. We have restricted ourselves to the domain
$\kappa_C\in(\s012,1)$ \cite{footnote}. For {\it general} regulator $r(x)$ 
we obtain  
\begin{eqnarray}\label{optresult} 
\kappa_C=0.59535...\,, \qquad \qquad \alpha_s= 2.9717... \,,
\end{eqnarray} 
which agrees with the analytic DS result \cite{Lerche:2002ep}. This
might seem surprising given the completely different representations.
However, the equivalence can be proven as follows: in the derivation
of \eq{optresult} we employed the truncation \eq{parG2} and
\eq{vertices} with the additional constraint $\delta Z \approx 0$
under the momentum trace in \eq{flow}.  Integrating the flow \eq{flow}
in this setting from $0$ to $k$ leads to
\begin{eqnarray} 
z_A\,p^2 x^{\kappa_A}\, \delta Z_A&=&\Tr \bigl[ G\,S^{(3)}\, G \,S^{(3)}
 - \s012\, G\,S^{(4)}\, G\bigr]_0^k\,,
\nonumber\\
z_C\,p^2 x^{\kappa_C}\,\delta Z_C&=& 
\Tr\,\bigl[ G\,S^{(3)}\, G \,S^{(3)}\bigr]_0^k\,, 
\label{G0}
\end{eqnarray} 
for the gluon and ghost two point function.  \Eq{G0} already implies
an UV renormalisation at $p^2=k^2$. This is reflected by \eq{genfixed}
being finite. Within the present truncation the renormalised DS
equations used in \cite{Lerche:2002ep} are equivalent to \eq{G0}.
Beyond this truncation the DS equation and the flow equation differ.
Apart from additional $t$-derivative terms, all vertices in \eq{G0}
get dressed as distinguished
from the DS equation, where both bare and dressed vertices are
present. Consequently, the flow equation \eq{flow} carries 
the correct RG properties if the truncation for $\Gamma_k$
does. In particular, equation \eq{G0} with the truncation \eq{parG2}
and \eq{vertices} allows for a fully self-consistent renormalisation.
Both the left and right hand sides of \eq{G0} transform the same way
under RG scalings, including the purely gluonic terms dropped in
\eq{genfixed}. Indeed, it is the latter fact which makes their
neglection fully consistent. The RG properties are most obvious for
regulators proportional to the wave function renormalisations $z$,
$R\simeq z\, \tilde R$. Then, the propagators can be written as
$G=z^{-1} G_0$ where $G_0$ does not involve further wave function
renormalisation factors.  Also, the vertex functions involve the
appropriate multiplicative powers of $z$ following from \eq{non-ren}
and \eq{vertices}.  \smallstep

We proceed by iterating the above solution, also taking into account
$\delta Z\neq 0$ on the right hand side of the flow.  Then
\eq{genfixed} can be used for a numerical iterative determination of
$\delta Z$ and $\kappa,\, \alpha_s$. We have computed $\alpha_s$ and
$\kappa$ for various classes of regulators. The exponent $\kappa_C$
takes values in an interval bounded by the values \eq{optresult} from
above and the iterated sharp cut-off result $\kappa_C\simeq0.539$ from
below. However, the values \eq{optresult} are singled out by their
physical relevance: $\kappa$ and $\alpha_s$ depend on the
regulator. This reflects the fact that the truncation
\eq{parG2},\eq{vertices} {\it and} the choice of $R$ define the
approximation to the full problem. We expect that global extrema of
$\kappa(R),\alpha_s(R)$ represent an optimised approximation based on
\eq{parG2},\eq{vertices}, see \cite{Litim:2001up}. In general it is
very difficult to translate this idea into properties of optimal
regulators or even in statements about the values of optimal
observables.  Here we present a new idea how to extract this
information: let us assume that we have identified a regulator $R_0$
which is an extremum in the space of regulators in the sense that
\begin{eqnarray}\label{extremum} 
\left[\Tr 
\,\bigl(\delta R \0{\delta}{\delta R}\bigr)\, \Gamma^{(n)}_k\right]_{R=R_0} =0 
\end{eqnarray} 
for all $\delta R$. \Eq{extremum} implies that $\kappa(R_0),\,
\alpha_s(R_0)$ are extrema, reflecting an optimisation of
physical quantities. 
Note that the $z_n(R_0)$ by themselves are not
physical quantities but they can be extrema up to RG scalings.
However, \eq{extremum} is more restrictive. For $n=2$, \eq{extremum}
is the additional demand, that $\delta Z(x)$ is an extremal curve.
Hence the question arises whether solutions $R_0$ to \eq{extremum}
exist at all: the fixed point assumption underlying the whole analysis
is valid for regulators $R$ that only lead to a suppression of
infrared modes. This implies a monotonous interpolation between the
physical infrared regime $ k^2\ll p^2\ll \Lambda^2_{\rm QCD}$ and the
infrared regularised regime $p^2\ll k^2$. Then the extremal curves
exist and are given by vanishing $\delta Z$ \cite{footnote3}. 
\smallstep

The above considerations allow us to identify the optimal values for
$\kappa_C(R)$ and $\alpha_s(R)$ without further computation. The correction
terms proportional to $\delta Z$ under the momentum trace are
negligible near optimal regulators and we are left with the
non-iterated equation \eq{genfixed} with an optimal regulator $R_0$.
However, we have already shown that the non-iterated solutions are
independent of the regulator and we conclude that
$\kappa_C(R_0),\alpha_s(R_0)$ are given by \eq{optresult}.  We are
also in a position to device optimal regulators. For these regulators
the effects of $\delta Z$ on the right hand side of \eq{genfixed} have
to disappear. They need not to satisfy \eq{extremum} which only
applies to a small sub-set of optimal regulators. In Fig.~1 we show
$\delta Z_{A/C}$ for the class of regulators
\begin{eqnarray}\label{classofregs}
r_C(x)=z_C\,\theta^{-1}(1-x)\,, \qquad  r_A(x)=\gamma\0{z_A}{ x(1+x)}\, . 
\end{eqnarray}
The $\gamma$-dependence of $\delta Z_A$ is negligible on the scale of
Fig.~1.  The ghost regulator in \eq{classofregs} is a sharp cut-off
while the gluon regulator is a rather soft mass-like cut-off. In the
limit $\gamma\to \infty$ the solution of \eq{genfixed} approaches the
optimal values \eq{optresult}. One verifies that all contributions from
$\delta Z$ to the right hand side of \eq{genfixed} tend to zero for
$\gamma\to \infty$ and we are left with the non-iterated result
\eq{optresult}.  Hence \eq{classofregs} with $\gamma\to \infty$
belongs to the class of optimal regulators. It is not in the extremal
subset as only the ghost propagator satisfies \eq{extremum} with
$\delta Z_C(x\neq 0)=0$. \smallstep

\begin{figure}[ht]
\begin{picture}(100,130)
\put(-50,-10){\epsfig{file=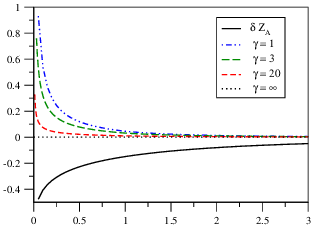,width=200pt}}
\put(155,3){{$x$}}\put(-67,100){{$\delta Z$}}
\put(155,3){{$x$}}\put(-5,70){{$\delta Z_C$}}
\end{picture}
\vskip.4cm
\begin{minipage}{.92\hsize}
 {\bf Figure 1:} $\delta Z$ for regulators with \eq{classofregs} 
\end{minipage} 
\end{figure}
\vskip-.3cm
 
In summary, we have shown how the non-trivial momentum dependence of
Green functions in the infrared sector is extracted within the
exact renormalisation group. To that end we have developed new fixed
point and optimisation techniques applicable to this situation.
This reasoning has been applied to the infrared behaviour of ghost and
gluon propagators in Landau gauge QCD. Our results provide further
evidence for the Kugo-Ojima scenario of confinement: the ghost
propagator is infrared enhanced while the gluon propagator develops a
mass gap.  In a truncation with dressed propagators, we have recovered
previous results from Dyson-Schwinger equations in a very simple
manner. This is noteworthy also in view of the qualitative differences
between these approaches.  Our method allows for a straightforward
improvement of the truncation by adding vertex corrections and further
terms with the correct RG properties. The same applies to dynamical
quarks. In both extensions the correct RG scaling as well
as the inherent finiteness of the integrated flow are most important.
Further results and details will be presented elsewhere.\smallstep

\noindent {\it Acknowledgements} \\[-2.5ex]

\noindent 
S.N.\ is supported by the DFG, contract SM70/1-1.


\begin{thebibliography}{99}
\bibitem{footnote} The
 $\kappa,\alpha_s$ run from their perturbative values into the regime
  $\kappa_C\in (\s012 ,1)$. We are interested in a more refined
  analysis of their infrared values rather than in the flow.
\bibitem{footnote3} This is valid up to RG scalings.
\end{thebibliography}
\end{document}